\documentclass[prx,aps,twocolumn,superscriptaddress]{revtex4}

\usepackage{dcolumn}
\usepackage{bm}
\usepackage{amssymb}
\usepackage{color}
\usepackage[pdftex]{graphicx}

\usepackage{amsmath}
\usepackage[linesnumbered,ruled]{algorithm2e}

\usepackage{url}
\usepackage[section]{placeins}
\usepackage[colorlinks=true,linkcolor=blue,citecolor=blue]{hyperref}

\begin{document}
\let\vec\mathbf

\title{Spin dynamics of the antiferromagnetic Heisenberg model on a kagome bilayer}

\author{Preetha Saha}
\affiliation{Department of Physics, University of Virginia, Charlottesville, VA 22904, USA}

\author{Depei Zhang}
\affiliation{Department of Physics, University of Virginia, Charlottesville, VA 22904, USA}

\author{Seung-Hun Lee}
\affiliation{Department of Physics, University of Virginia, Charlottesville, VA 22904, USA}

\author{Gia-Wei Chern}
\affiliation{Department of Physics, University of Virginia, Charlottesville, VA 22904, USA}

\date{\today}

\begin{abstract}
We study the spin dynamics of classical Heisenberg antiferromagnet with nearest neighbor interactions on a quasi-two-dimensional kagome bilayer. This geometrically frustrated lattice consists of two kagome layers connected by a triangular-lattice linking layer. By combining Monte Carlo with precessional spin dynamics simulations, we compute the dynamical structure factor of the classical spin liquid in kagome bilayer and investigate the thermal and dilution effects. While the low frequency and long wavelength dynamics of the cooperative paramagnetic phase is dominated by spin diffusion, weak magnon excitations persist at higher energies, giving rise the half moon pattern in the dynamical structure factor. In the presence of spin vacancies, the dynamical properties of the diluted system can be understood within the two population picture. The spin diffusion of the ``correlated" spin clusters is mainly driven by the zero-energy weather-van modes, giving rise to an autocorrelation function that decays exponentially with time. On the other hand, the diffusive dynamics of the quasi-free ``orphan" spins leads to a distinctive longer time power-law tail in the autocorrelation function. We discuss the implications of our work for the glassy behaviors observed in the archetypal frustrated magnet SrCr$_{9p}$Ga$_{12-9p}$O$_{19}$ (SCGO).
\end{abstract}

\maketitle


\section{Introduction}

The SCGO is one of the most intensely studied frustrated magnets~\cite{obradors88,ramirez90,ramirez92,broholm90,uemura94,lee96a,schiffer96,lee96b,ramirez00,keren00,mendels00,bono05,iida12,yang15}. Thermodynamically, SCGO does not exhibit any signs of magnetic ordering down to temperatures $T_g =$ 3.5--7 K, depending weakly on the vacancy concentration $x = 1-p$. Below $T_g$, the magnet enters an unconventional spin-glass phase. A cooperative paramagnetic regime, also known as a classical spin liquid, emerges at temperatures below the Curie-Weiss constant $\Theta_{\rm CW} \approx 500$~K.
Geometrically, SCGO belongs to a class of frustrated Heisenberg antiferromagnets on the so-called bi-simplex lattices~\cite{moessner98a,moessner98b,henley01,canals98}. These are networks of corner-sharing simplexes such as triangles and tetrahedra. Canonical examples include the pyrochlore~\cite{moessner98a,moessner98b} and kagome~\cite{chalker92,garanin99,henley09,chern13} antiferromagnets. In SCGO, the Cr$^{3+}$ ions with spin $S=3/2$ reside on a two-dimensional lattice consisting of corner-sharing tetrahedra and triangles, known as the kagome bilayer or pyrochlore slab, as shown in Fig.~\ref{fig:kagome}.
The strong short-range spin correlations in the low-temperature liquid phase result from the constraints of zero total spin in every simplex, a condition that minimizes the nearest-neighbor exchange interactions on such unit. 

Considerable experimental efforts have been devoted to understanding the unusual spin glass phase in SCGO~\cite{ramirez90,ramirez92,broholm90,uemura94,lee96b,keren00,mendels00,bono05,yang15}. Despite the characteristic field-cooled and zero-field-cooled hysteresis in the bulk susceptibility, several dynamical properties of its glassy phase are distinctly different from those of conventional spin glasses. These include the quadratic $T^2$ behavior of the specific heat~\cite{ramirez90,ramirez92},  the linear $\omega$-dependent dynamical susceptibility~$\chi''$~\cite{lee96b}, and a significantly weaker memory effect~\cite{samarakoon16}. Taken together, these features suggest that SCGO belongs to a new state of glassy magnets, dubbed the spin jam~\cite{yang15,samarakoon16}, that include several other magnetic compounds~\cite{yang16,samarakoon17}. The source of this unusual dynamical phase in SCGO, however, remains to be clarified. One plausible scenario is that quantum fluctuations transform the macroscopic degeneracy associated with the classical spin liquid of the kagome bilayer into the rugged energy landscape of spin jam~\cite{yang15,klich14}. It remains to be shown how the unusual glassy behaviors of the spin jam evolve from the spin dynamics of the cooperative paramagnet.

Toward this goal, we present in this paper the first systematic study of the dynamical properties of the bilayer-kagome classical spin liquid. By combining Monte Carlo simulations with energy-conserving Landau-Lifshitz dynamics, we compute the dynamical structure factor of the liquid regime. At the energy scales of the exchange interaction, we find signals of spin-wave excitations in the form of half moon pattern, replacing the pinch-point singularity of the static structure factor. On the other hand, the low-energy dynamics is dominated by spin-diffusion driven mostly by the zero-energy modes. The diffusion constant is found to depend weakly on temperature, but decrease significantly with increasing vacancy densities. 

Our results will also serve as an important benchmark against which dynamical behaviors induced by other perturbations can be compared. Of particular interest are those perturbations, such as quantum order-by-disorder, that give rise to glassy dynamics characteristic of either the conventional spin-glass or the exotic spin-jam states. 
It is also worth noting that the dynamical properties of classical spin liquid has been extensively studied for Heisenberg antiferromagnets on both pyrochlore~\cite{moessner98a,moessner98b,conlon09} and kagome lattices~\cite{keren94,robert08,taillefumier14,bilitewski19}. 
Another aim of this paper is thus to compare the spin dynamics of bilayer kagome against these two well studied bi-simplex frustrated magnets.

The rest of the paper is organized as follows. In section~\ref{sec:model}, we discuss the ground-state manifold of Heisenberg antiferromagnet on the kagome bilayer. We also outline the numerical framework that combines Monte Carlo simulation with energy-preserving Landau-Lifshitz dynamics method for computing the dynamical structure factor of a classical spin liquid. Magnetic excitations revealed from the dynamical structure factor are discussed in Sec.~\ref{sec:magnon}. In particular, half-moon features, which are the dynamical manifestation of the famous pinch-point structure at finite energies, are highlighted. Systematic analysis of the low-energy spin dynamics, which is dominated by diffusive modes, is presented in Sec.~\ref{sec:diffusion}. We present in Sec.~\ref{sec:dilution} dynamical features due to quenched disorder introduced by vacancies. Of particular interest is the emergence of quasi-free orphan spins that interact with each other through a week effective interaction mediated by the background spin-liquid. We conclude in Sec.~\ref{sec:discussion} with a brief summary and outlook on future work.

\section{Model and method}

\label{sec:model}

\begin{figure}
\includegraphics[width=.95\columnwidth]{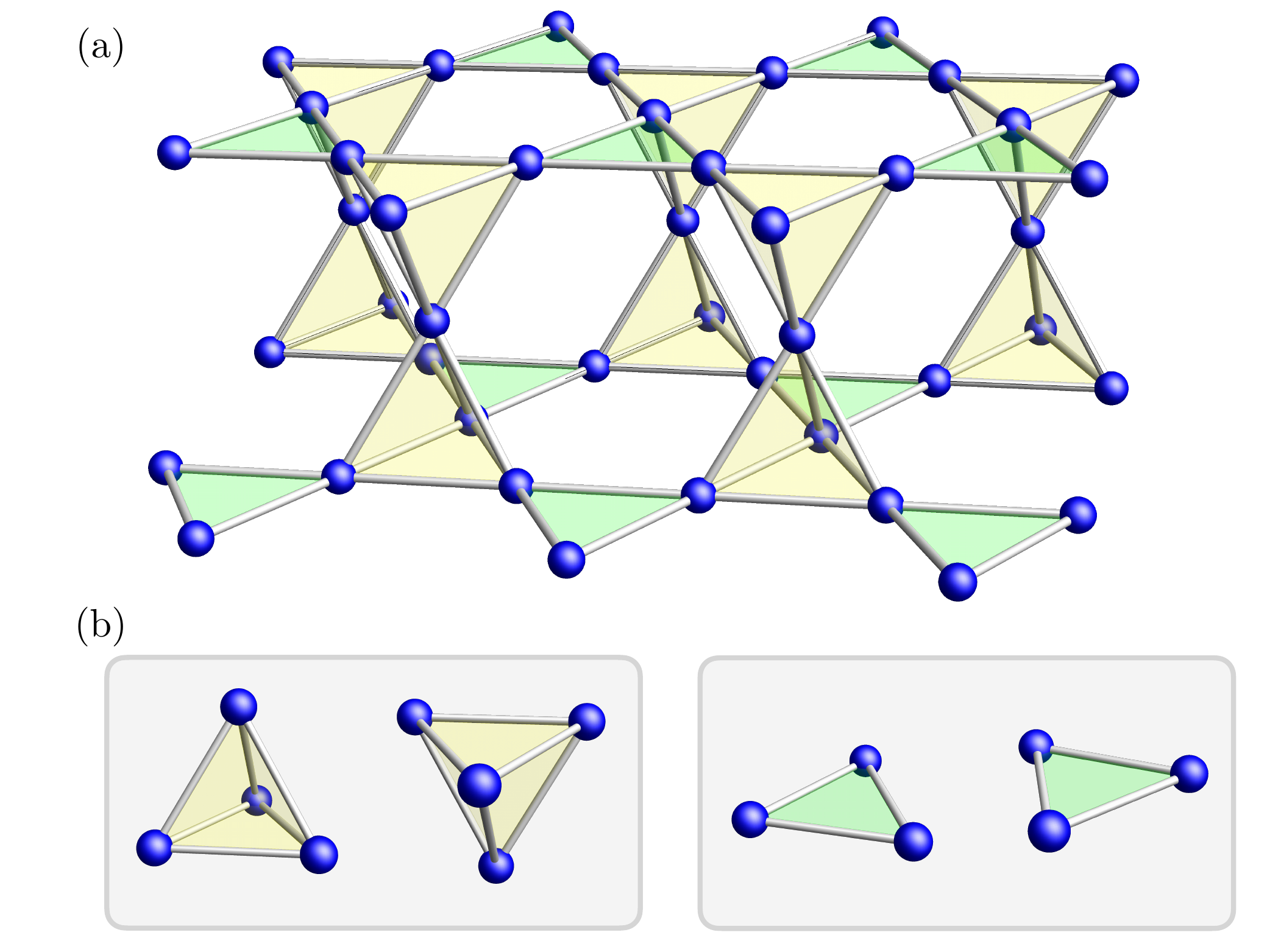}
\caption{
\label{fig:kagome} (Color online) Top: Lattice structure of kagome bilayer. It can be viewed as a quasi-two-dimensional network of corner-sharing simplexes. There are two kinds of simplexes: tetrahedron and triangles, both comes with two (opposite) orientations, as shown in the bottom panel. Spins in the ground states satisfy the constraint that total spin in both types of simplexes is zero: $\mathbf L_{\boxtimes} = \mathbf L_{\triangle} = 0$.
}
\end{figure}

We consider the classical Heisenberg model with nearest neighbor interactions on the kagome bilayer
\begin{eqnarray}
	\label{eq:Hamil}
	\mathcal{H} &=&J \sum_{\langle i j \rangle} \mathbf {S}_{i}  \cdot \mathbf {S}_{j} 
\end{eqnarray}
Here $J > 0$ is the antiferromagnetic exchange, $\langle ij \rangle$ denotes nearest-neighbor pairs, and the classical spins $\mathbf {S}_{i}$ are unit vectors.
The kagome bilayer has a Bravais triangular lattice with a unit cell consisting of two corner sharing tetrahedra of opposite orientation. The bases of the tetrahedra in the two kagome layers are connected by triangle units; see Fig.~\ref{fig:kagome}. The triangle and tetrahedron are the regular simplexes with $q=3$ and $q=4$ corners, respectively. Importantly, because of this corner-sharing simplex structure, the exchange interaction can also be expressed as a sum of the squared total spin of both types of simplexes
\begin{eqnarray}
	\mathcal{H} = \frac{J}{2}\sum_{\boxtimes}  \mathbf {\vec{L}}_{\boxtimes}^2 + \sum_{\triangle} \mathbf L_{\triangle}^2   +  \mbox{const.}  
\end{eqnarray}
Here $\mathbf L_\boxtimes =  \sum_{i\in \boxtimes}\vec{S}_{i}$ denotes total spin in a tetrahedron, $\mathbf{L}_{\triangle} = \sum_{i\in\triangle} \mathbf S_i$ denotes total spins of a triangle, and $\sum_{\boxtimes}$ and $\sum_{\triangle}$ indicate summation over tetrahedra and triangles, respectively, in the kagome-bilayer lattice. One can immediately see that the exchange energy is minimized by the condition that total spin of every simplexes is zero: 
\begin{eqnarray}
	\label{eq:gs}
	\mathbf L_\boxtimes = \mathbf L_\triangle =  0,
\end{eqnarray} 
The ground-state condition is confirmed by our Monte Carlo simulations. The fact that a macroscopic number of spin configurations satisfy the minimum energy condition leads to a classical spin liquid regime at temperatures $T \lesssim J$. Indeed, our Monte Carlo simulations show no signs of phase transition down to temperatures $T \approx 0.001 J$, consistent with previous studies~\cite{arimori01,kawamura02,sen11,sen12}. Instead, a spin-disordered phase with strong short-range correlation is obtained at low temperatures.


In general, there are two types of spin dynamics in the liquid regime. At short time scales, or high frequencies ($\omega \sim J$), there are spin-wave excitations corresponding to small amplitude deviations from the ground-state manifold. These excitations are similar to the magnons in unfrustrated magnets. On the other hand, the macroscopic number of zero modes, or the weather-vane modes, that connect different ground states dominate the long-time dynamical behaviors of the frustrated bi-simplex antiferromagnet. The resultant drifting of the system in the ground-state manifold gives rise to spin-diffusion behaviors and an exponential decaying spin autocorrelation. In the following, we discuss our simulation results within this general picture.

The equation of motion for classical spins is given by the Landau-Lifshitz equation
\begin{eqnarray}
	\frac{d\vec{S}_{i}}{dt} = -\vec{S}_{i} \times \frac{\partial \mathcal{H}}{\partial \mathbf S_i} = - J \sum_{j}\!' \, \mathbf S_i \times \mathbf S_j,   
\label{eq:LL}
\end{eqnarray}
where the prime indicates summation is restricted to the nearest neighbors of the $i$-th spin.
Here we numerically integrate the Landau-Lifshitz equation to compute the dynamical structure factor of the classical spin liquid. Low temperature Monte Carlo simulations are first used to obtain spin configurations in equilibrium of a specified temperature. These are then used as the initial states for the energy-conserving precession dynamics simulations. An efficient semi-implicit integration algorithm~\cite{mentink10} is employed to integrate the above Landau-Lifshitz equation. The high efficiency of the algorithm comes from fact that it preserves the spin length at every time step and the energy values are well conserved with time irrespective of the step size or the time span of the simulation. From the numerically obtained spin trajectories $\mathbf S_i(t)$, we compute the dynamical correlation function $\mathcal{S}(\mathbf q, t)$ 
\begin{eqnarray}
	\mathcal{S}(\mathbf q, t) = \langle \mathbf S_{\mathbf q}(t) \cdot \mathbf S_{\mathbf q}^*(0) \rangle,
\end{eqnarray}
where $\mathbf S_{\mathbf q}(t) \equiv \sum_i \mathbf S_i(t) \exp(i \mathbf q \cdot \mathbf r_i) / \sqrt{N}$ is the spatial Fourier transform of the instantaneous spin configuration, and $\langle \cdots \rangle$ denotes the ensemble average over independent initial states of a given temperature. The dynamical structure factor is then given by
\begin{eqnarray}
	\mathcal{S}(\mathbf q, \omega) &=& \int \mathcal{S}(\mathbf q, t) e^{-i \omega t} dt \nonumber \\
	& = & \frac{1}{N}\sum_{ij} \int dt \langle \mathbf S_i(t) \cdot \mathbf S_j(0) \rangle e^{-i \omega t} dt,
\end{eqnarray}
which is essentially the space-time Fourier transform of the spin-spin correlator $C_{ij}(t) \equiv \langle \mathbf S_i(t) \cdot \mathbf S_j(0) \rangle$.

\begin{figure}[t]
\includegraphics[width=.99\columnwidth]{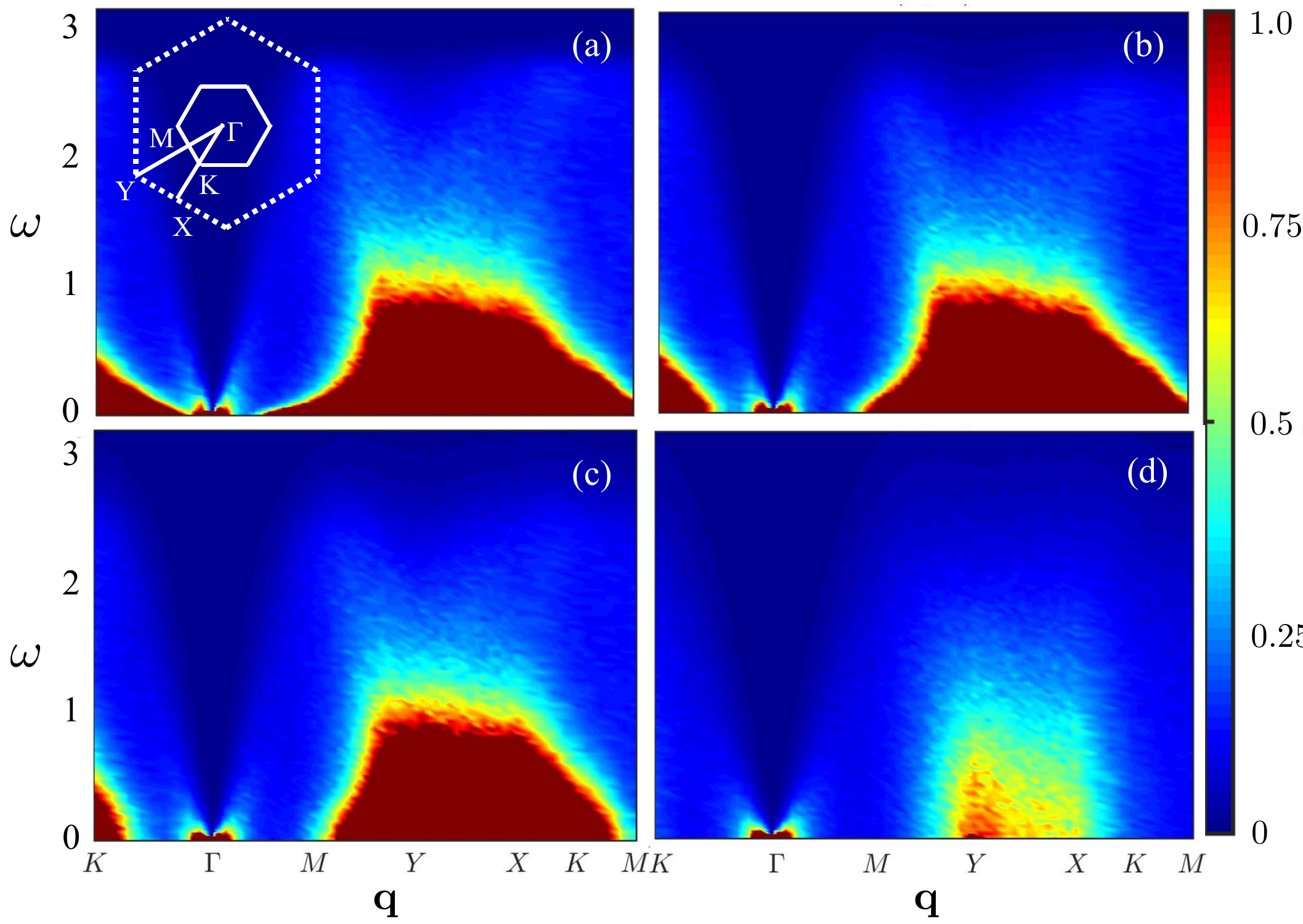}
\caption{   
\label{fig:dyn_sf} (Color online) Temperature-scaled dynamical structure factor $\beta\,\mathcal{S}(\mathbf q, \omega)$ of the classical spin liquid in the bilayer kagome antiferromagnet at four different temperatures:  (a) $T/J=0.01$, (b) 0.05, (c) 0.1, and (d)~0.6; here $\beta = 1/T$. The linear size of the simulated lattice is $L = 30$, with number of spins $N = 7\times L^2$.
}
\end{figure}

\section{Magnons and half moon patterns}

\label{sec:magnon}

\begin{figure}
\includegraphics[width=.99\columnwidth]{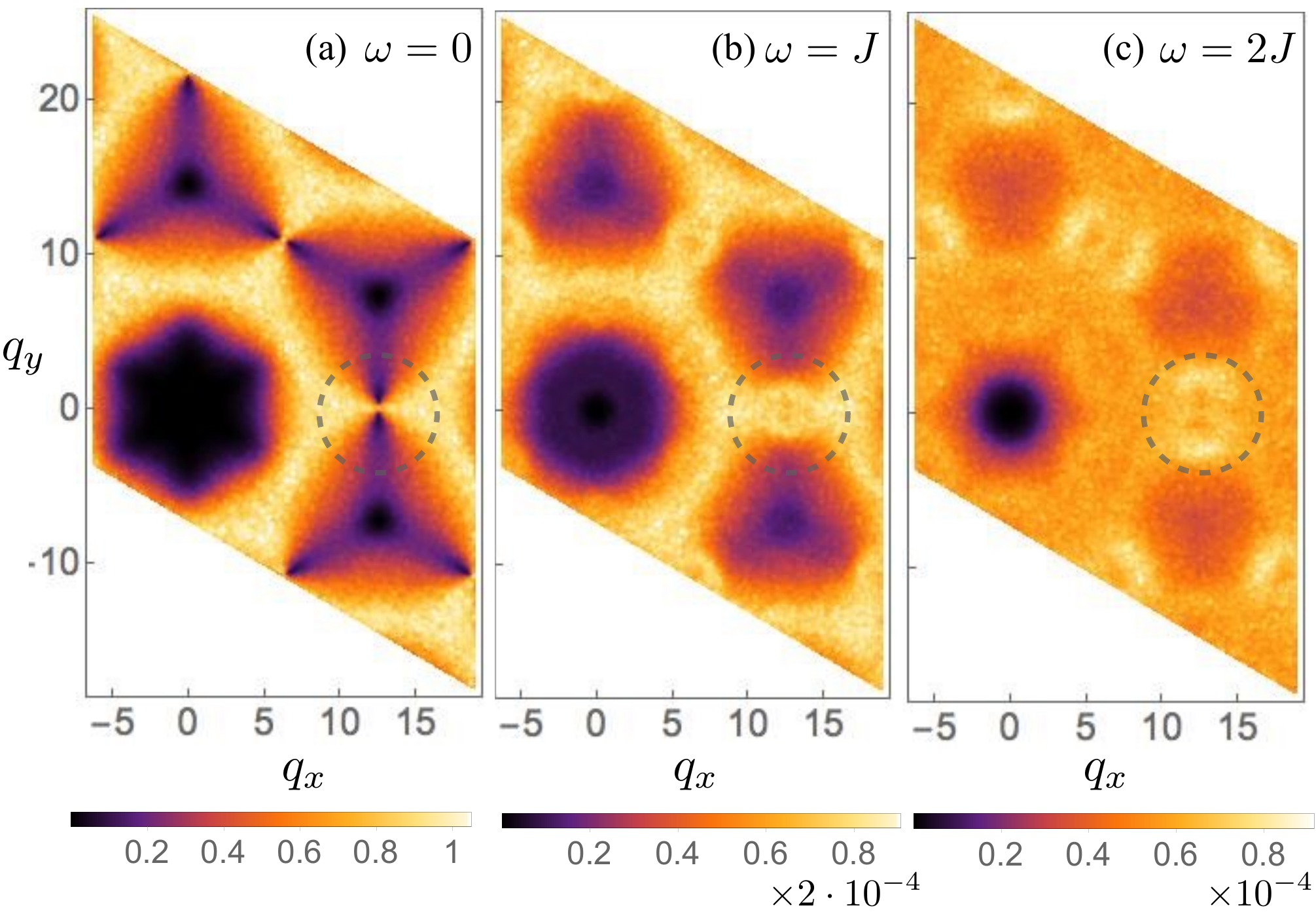}
\caption{(Color online)  
\label{fig:dsf} 
(Color online) Density plots of the dynamical structure factor $\mathcal{S}(\mathbf q, \omega)$ at $T = 0.005\,J$ in the reciprocal space: (a) $\omega = 0$, (b) $\omega = J$, and (c) $2 J$. The system size is $L = 30$. The dashed circles indicate the pinch point at $\omega = 0$, and the half moon pattern at higher energies. 
}
\end{figure}

The intensity plot of the scaled dynamical structure factor $\beta\,\mathcal{S}(\mathbf q, \omega)$, where $\beta = 1/T$, is shown in Fig.~\ref{fig:dyn_sf} for four different temperatures. The spin excitations are clearly dominated by the low-energy quasi-static fluctuations that extend over most of the Brillouin zone; also see Fig.~\ref{fig:dsf} for the density plots of $\mathcal{S}(\mathbf q, \omega)$ in the reciprocal space at constant energies. At low temperatures, the similar patterns of the quasi-static excitations indicating nontrivial scaling behaviors to be discussed below.
Moreover, the relatively weak excitations at higher energies $\omega \gtrsim J$ result from the magnon fluctuations in the vicinity of an instantaneous ground state. Contrary to the kagome antiferromagnets~\cite{robert08,taillefumier14}, no sharp propagating modes can be seen in the dynamical structure factor of the bilayer kagome.

The static structure factor, corresponding to Fig.~\ref{fig:dsf}(a) with $\omega = 0$, exhibits sharp pinch points which are a hallmark of highly correlated spin liquid in bi-simplex frustrated magnets. The source of these singularities can be attributed to the ground-state constraints Eq.~(\ref{eq:gs}), which translates into a solenoid condition $\nabla \cdot \mathbf B = 0$ for an emergent ``magnetic" or flux field that is a coarse-grained representation of the spin configuration. This in turn gives rise to an anisotropic dipolar-like correlation of the flux field, which manifests itself as the pinch-point singularity in the reciprocal space~\cite{garanin99,isakov04,henley05}.

At finite temperatures, the width of the pinch-point is roughly proportional to $\sqrt{T}$~\cite{conlon10}. Interestingly, the pinch point is also smeared with increasing $\omega$, and is replaced by the so-called half moon pattern at $\omega \gtrsim J$, as shown in Fig.~\ref{fig:dsf} (b) and (c). Similar features, called the ``excitation rings" have been observed in the finite-energy dynamical structure factor of the coplanar spin liquid phase of kagome~\cite{robert08}. It has been pointed out that the half-moon can be viewed as the pinch-point with a dispersive dynamical flux field~\cite{yan18}. These crescent patterns at high energies are the remnants of the propagating magnons mentioned above. Compared with the coplanar phase in kagome, the half-moon feature is much weaker in the liquid phase of bilayer kagome, indicating less rigid local structures in the instantaneous ground state.

\section{Spin diffusion}
\label{sec:diffusion}

\begin{figure}
\includegraphics[width=0.99\columnwidth]{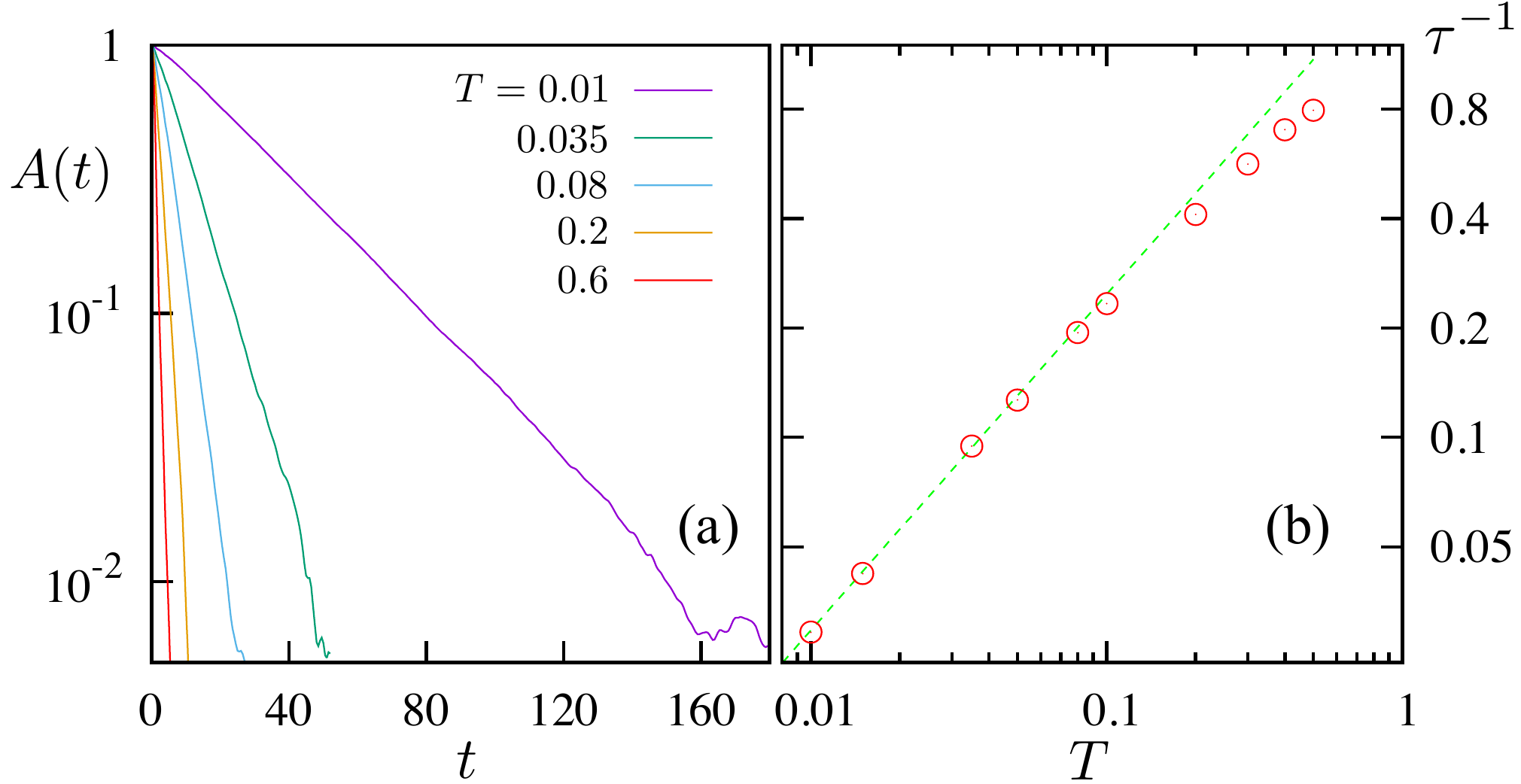}
\caption{  
\label{fig:At} (Color online) (a) The ensemble averaged spin autocorrelation function $A(t) = \sum_i \langle \mathbf S_i(t) \cdot \mathbf S_i(0) \rangle / N$ on a $L = 30$ lattice for varying temperatures. (b) Extracted relaxation time $\tau$ of $A(t) = \exp(-t/\tau)$ as a function of temperature. The dashed line shows the power-law $\tau \sim T^{-0.924}$ dependence. 
}
\end{figure}

The relatively weak half moon excitations also indicate a dominating spin diffusive dynamics in bilayer kagome. In general, spin diffusion dominates the excitation spectrum of disordered Heisenberg systems in the hydrodynamic limit~\cite{halperin69,muller88}. In frustrated magnets, this diffusion results from the macroscopic number of zero-energy modes in the instantaneous ground state, causing the system to wander around the degenerate manifold. One particular manifestation of this diffusion is the decay of the spin autocorrelation function
\begin{eqnarray}
	\label{eq:At}
	A(t) = \frac{1}{N} \sum_i \langle \mathbf S_i(t) \cdot \mathbf S_i(0) \rangle  = \sum_{\mathbf q} \mathcal{S}(\mathbf q, t), 
\end{eqnarray}
where again $\langle \cdots \rangle$ is the thermal average, which is achieved through averaging over independent initial states from Monte Carlo simulations. Fig.~\ref{fig:At}(a) shows $A(t)$ as a function of time for various temperatures obtained from a $L=30$ system. The decay of the autocorrelation function is found to be exponential $A(t) \sim \exp(-t/\tau)$ in the low temperature regime, and the numerically extracted time constant $\tau$ is shown in Fig.~\ref{fig:At}(b) as a function of temperature.


The nearly linear segment in the log-log plot suggests a power-law dependence $\tau \sim T^{-\zeta}$, where the numerically obtained exponent $\zeta = 0.924 \pm 0.015$, which is close to~1 as predicted by a soft-spin Langevin dynamics model for frustrated magnets with macroscopic ground-state degeneracy~\cite{conlon09}. The exponential decay with $\tau \sim 1/T$ is consistent with the zero-mode driven spin-diffusion scenario~\cite{moessner98a,conlon09}, since the zero modes have no intrinsic energy scales, and the only relevant one is set by the inverse temperature. This result is also in stark contrast to the high-$T$ conventional paramagnet in which the spin-diffusion is shown to produce a power-law tail in the autocorrelation function~\cite{muller88,gerling90,gerling90b,constantoudis97,bagchi13}.

\begin{figure}[t]
\includegraphics[width=0.99\columnwidth]{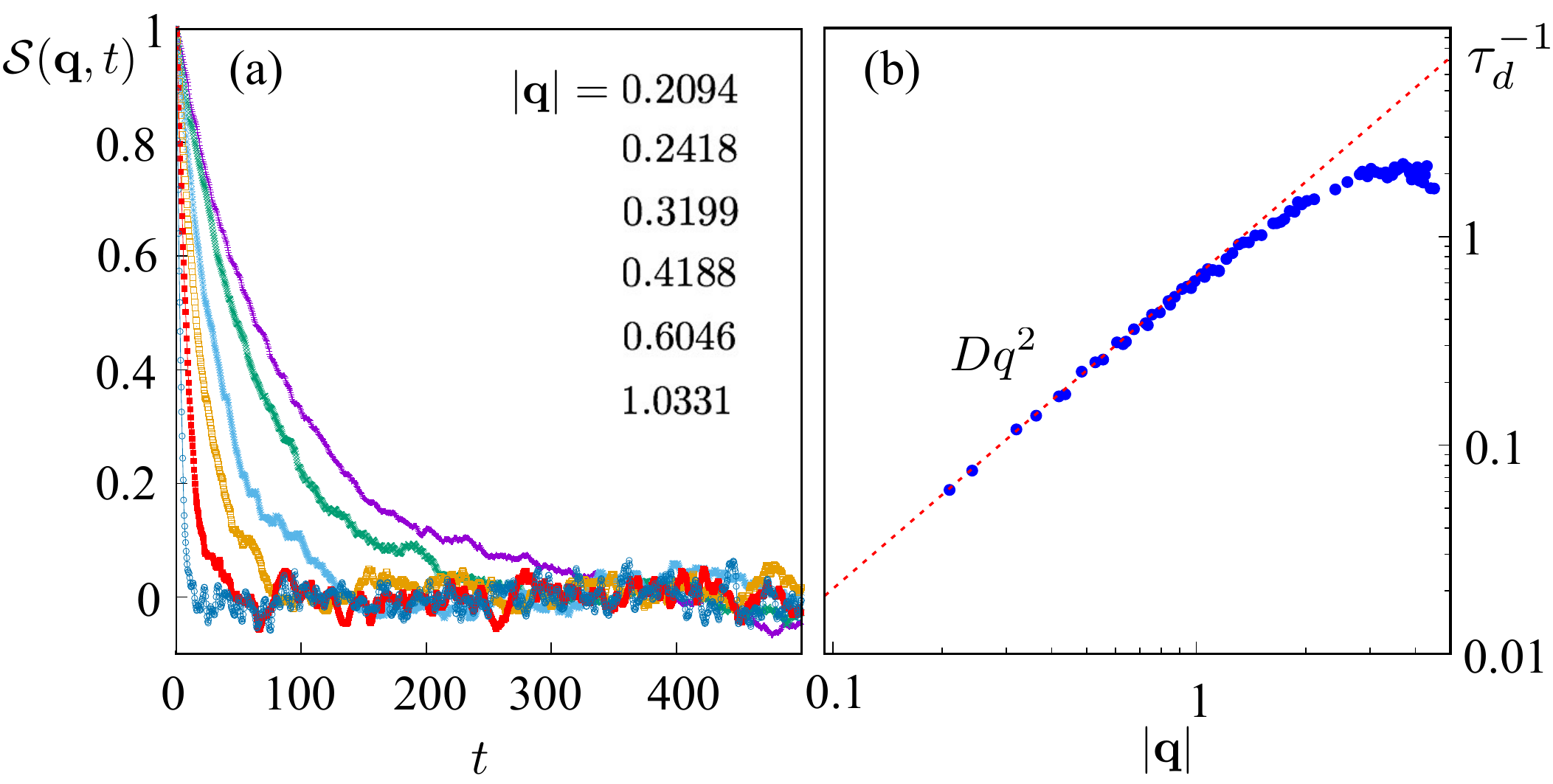}
\caption{
\label{fig:QAt} (Color online) (a) Time dependence of the normalized dynamical correlation function $\mathcal{S}(\mathbf{q},t) / \mathcal{S}(\mathbf q, 0)$ at $T = 0.5J$ for various wave vectors close to the zone center. The simulated system size is $L = 60$. (b) The inverse relaxation time $\tau_d^{-1}$ extracted from panel (a) as a function of $|\mathbf q|$. For each wave vector the data is fitted to an exponentially decaying function. 
}
\end{figure}

While the microscopic mechanisms of spin-diffusion could be thermal or quantum fluctuations, or the large number of zero modes in frustrated systems, fundamentally the diffusive spin dynamics is related to the fact that the total spin density $\mathbf m = \sum_i \mathbf S_i / N$ is a constant of the equation of motion. By combining the continuity equation $\partial \mathbf m/ \partial t + \nabla \cdot \mathbf j = 0$ with a phenomenological Fick's law for local spin current $\mathbf j = -D \nabla \mathbf m$, one arrives at the familiar diffusion equation for the magnetization density. 
In the hydrodynamic regime, this introduces a diffusion timescale $\tau_d = 1/D q^2$ for perturbations characterized by wave vector $\mathbf q$. This is indeed confirmed by our dynamical simulations. Fig.~\ref{fig:QAt}(a) shows the time dependence of the dynamical correlation function $\mathcal{S}(\mathbf q, t)$ for various wave vectors. Each curve is obtained after averaging over 500 independent initial states from Monte Carlo simulations. The correlation function is found to decay exponentially with time: $\mathcal{S}(\mathbf q, t) \sim \exp(-t / \tau_d)$, where the numerically extracted relaxation time, shown in Fig.~\ref{fig:QAt}(b), is isotropic in the reciprocal space and follows nicely the expected behavior $\tau_d^{-1} = D q^2$ for wave vectors close to the Brillouin zone center.

\begin{figure}
\includegraphics[width=0.94\columnwidth]{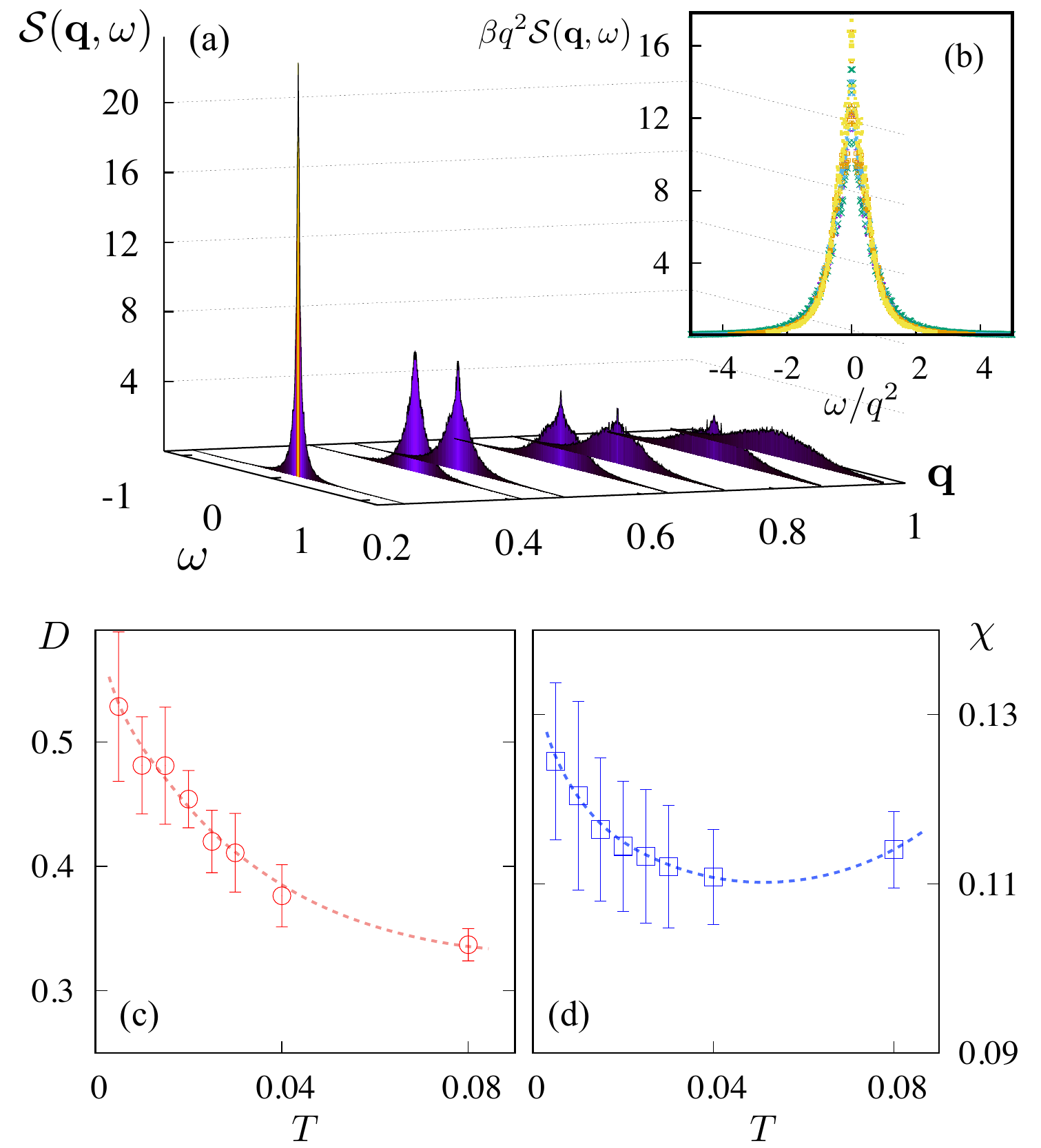}
\caption{
\label{fig:Sqw} (Color online) (a) Dynamical structure factor $\mathcal{S}(\mathbf q, \omega)$ as a function of $\omega$ at varying wave vectors~$\mathbf q$ at a temperature $T = 0.01J$.  (b) Scaling collapse according to  Eq.~(\ref{eq:sqw}) for data points from different wave vectors shown in panel (a). These curves are well approximated by a Lorentzian centered on $\omega = 0$. By fitting the collapsed data-points to the scaling function, the numerically extracted spin diffusion coefficient $D$ and static susceptibility $\chi$ (normalized to the value at $T =0.01$) are shown in panels (c) and (d), respectively, as functions of temperature.
}
\end{figure}

More generally, here we try to understand our results using the hydrodynamic theory of the paramagnetic state, which suggests a generalized dynamical susceptibility: $\chi(\mathbf q, \omega) = -\chi(\mathbf q) \, {D q^2}/(Dq^2 -i \omega)$~\cite{foster75,halperin69}, where $\chi(\mathbf q)$ is the static susceptibility at wave vector $\mathbf q$ and $D$ is the spin diffusion coefficient. The dynamical structural factor is obtained through the fluctuation-dissipation theorem: $\mathcal{S}(\mathbf q, \omega) \approx 2 [n_B(\omega) + 1]\, {\rm Im}\chi(\mathbf q, \omega)$, where $n_B(\omega) = 1/(e^{\beta\omega}-1)$. In the $\omega \ll T$ regime, assuming $\chi(\mathbf q) \approx \chi$ is a constant for small $q$, the dynamical structure factor can be expressed in a scaling form
\begin{equation} 
	  \beta q^2 \mathcal{S}(\vec{q},\omega) =  \chi \frac{2D}{(\omega/q^2)^2 +D^2},
	\label{eq:sqw}
\end{equation}
A similar result can be obtained from the Langevin soft-spin model~\cite{conlon09}. By plotting $\beta q^2 \mathcal{S}$ versus $\omega/q^2$, we find nice data collapsing from curves of different wave vectors, as shown in Fig.~\ref{fig:Sqw} (a) and (b), indicating a static susceptibility that indeed weakly depends on $\mathbf q$ for wave vectors close to zone center. On the other hand, we find that the collapsing of data points from different temperatures is not very satisfactory. Instead, we fit the collapsed data points from each temperature with the Lorentzian scaling function in Eq.~(\ref{eq:sqw}) and extract both the spin diffusion coefficient $D$ and static susceptibility $\chi$. The temperature dependence of these two quantities are shown in Fig.~\ref{fig:Sqw}(c) and (d).  The spin-diffusion coefficient decreases quite appreciably with temperature, while the susceptibility remains roughly the same within the error bars.

\section{Dilution effects}
\label{sec:dilution}

\begin{figure}
\includegraphics[width=.99\columnwidth]{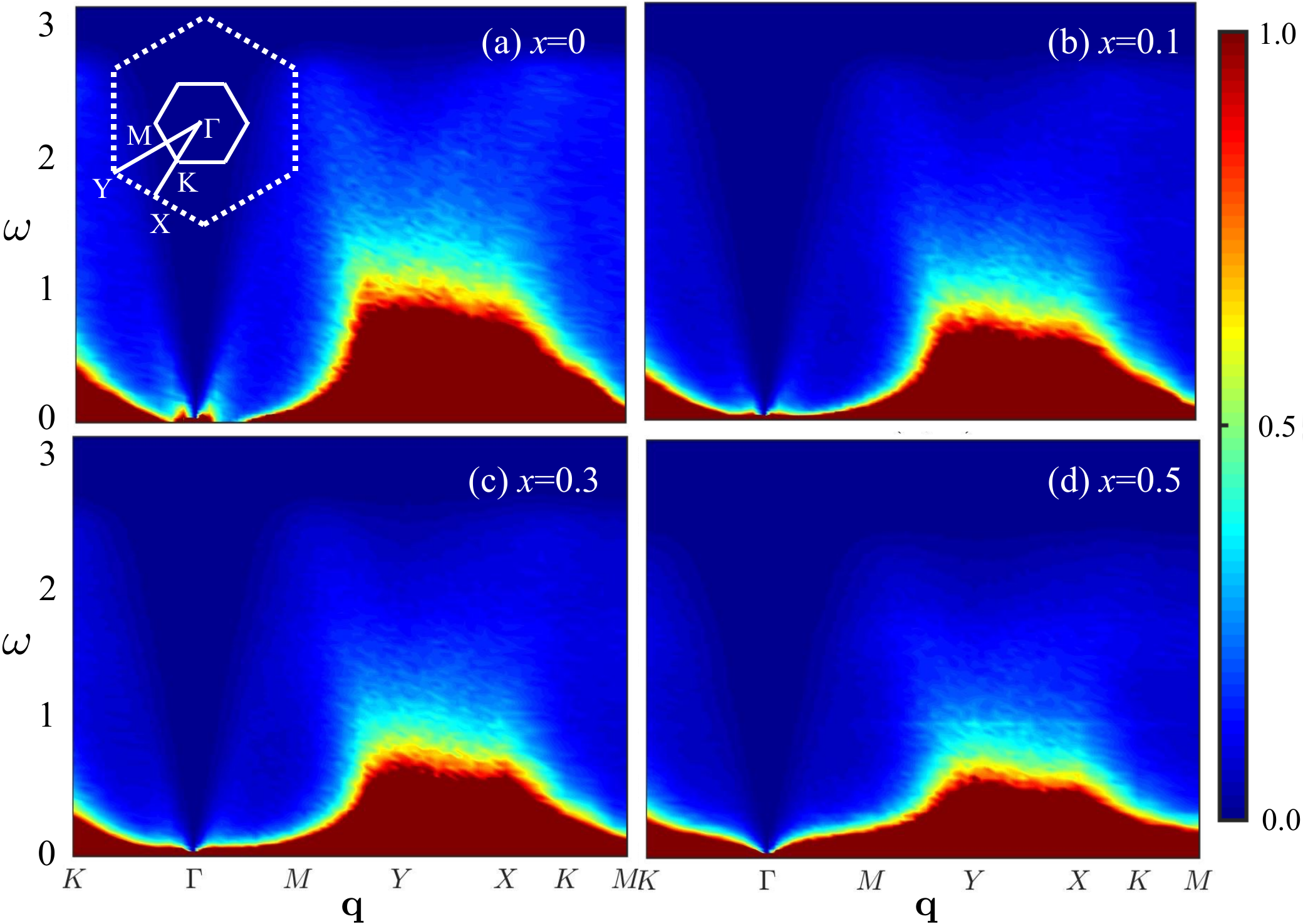}
\caption{   
\label{fig:dsf_vacancy} 
(Color online) Dynamical structure factor $\mathcal{S}(\mathbf q, \omega)$ of the classical spin liquid in the diluted bilayer kagome antiferromagnet at $T = 0.01$ for four different vacancy concentrations:  (a) $x=0$ (no dilution), (b) 0.1, (c) 0.3, and (d)~0.5.  The linear size of the simulated lattice is $L = 30$. The density plots for diluted systems ($x \neq 0$) are further averaged over 50 different disorder configurations.
}
\end{figure}

We next investigate the effect of dilution on the spin dynamics of the liquid phase. Previous studies have indicated that dilution with non-magnetic vacancies does not induce the spin-glass behavior of SCGO~\cite{henley01,shender93}. In fact, the condition Eq.~(\ref{eq:gs}) is satisfied for every simplex, for both tetrahedron and trinagle, in the ground states even for strong dilution~\cite{henley01}. Consequently, a macroscopic degeneracy remains and the low-$T$ phase seems well approximated by a Coulombic classical spin liquid. 
To demonstrate this, we compute the dynamical structure factor of the diluted kagome bilayer using a combination of Monte Carlo simulations with the energy-conserving Landau-Lifshitz dynamics simulations. Fig.~\ref{fig:dsf_vacancy} shows the computed $\mathcal{S}(\mathbf q, \omega)$ at $T = 0.01J$ for four different vacancy concentrations.  In addition to the thermal average over independent initial states, the $\mathcal{S}(\mathbf q, \omega)$ of the diluted system is computed with a further average over the disorder, or different vacancy configurations. Interestingly, we find no dramatic change to the calculated $\mathcal{S}(\mathbf q, \omega)$ even for vacancy density as high as $x = 0.5$. The quasi-static excitations show similar patterns for all concentrations, although both the energy of spin-wave-like excitations at large $\omega$ and the bandwidth of the quasi-static excitations are slightly reduced with increasing vacancy concentrations.

\begin{figure}
\includegraphics[width=.95\columnwidth]{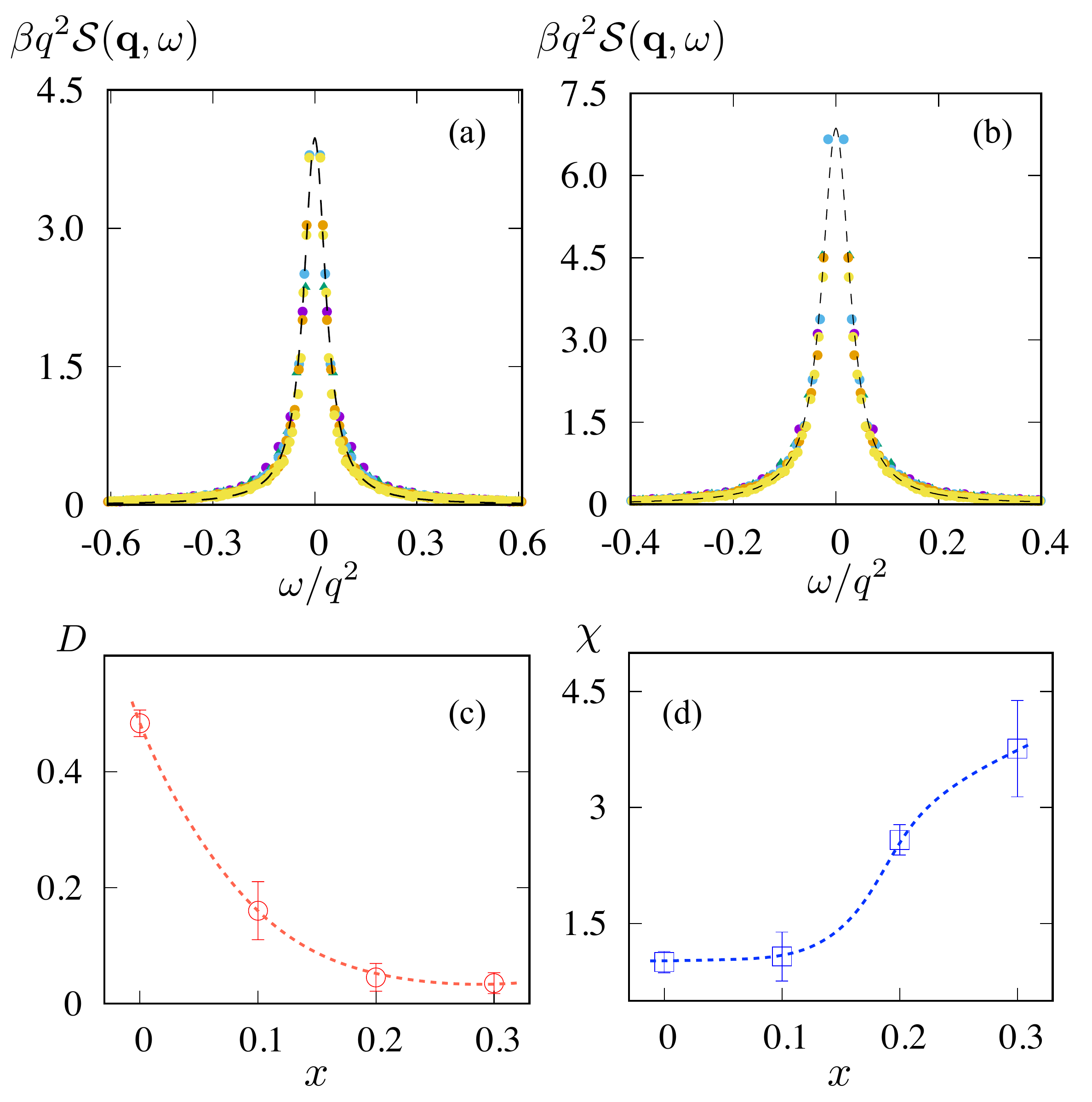}
\caption{
\label{fig:vacancy} (Color online) Data points collapsing of the scaled dynamical structure factor $\beta q^2 \mathcal{S}$ versus $\omega/q^2$ for diluted bilayer kagome with (a) $x = 0.2$ and (b) $x = 0.3$ vacancy concentrations at a temperature $T = 0.01J$. The dashed lines correspond to the Lorentzian scaling function in Eq.~(\ref{eq:sqw}). The extracted diffusion coefficient $D$ and static susceptibility $\chi$ (normalized with respect to $x = 0$) as functions of the vacancy concentration are shown in panels (c) and (d), respectively. 
}
\end{figure}

Focusing on the small-$\omega$ and $\mathbf q$ regime, we found that the dynamical structure factor is still well approximated by the scaling function of Eq.~(\ref{eq:sqw}), as shown in Fig.~\ref{fig:vacancy}(a) and (b), indicating a dominating spin diffusion behavior. The diffusion coefficient $D$ extracted from the data-point collapsing is plotted in Fig.~\ref{fig:vacancy}(c) as a function of~$x$. The reduced diffusivity with increasing vacancy concentration indicates a longer relaxation time $\tau_d = 1/D q^2$, or a slower dynamics, caused by the disorder, although the system remains liquid-like. Fig.~\ref{fig:vacancy}(d) shows the extracted static susceptibility $\chi$ in the $q \to 0$ limit versus vacancy concentration $x$. This trend is consistent with the two population picture since the quasi-free orphan spins dominate the low-$T$ static susceptibility, hence $\chi$ increases with the vacancy concentration.

\begin{figure}
\includegraphics[width=.95\columnwidth]{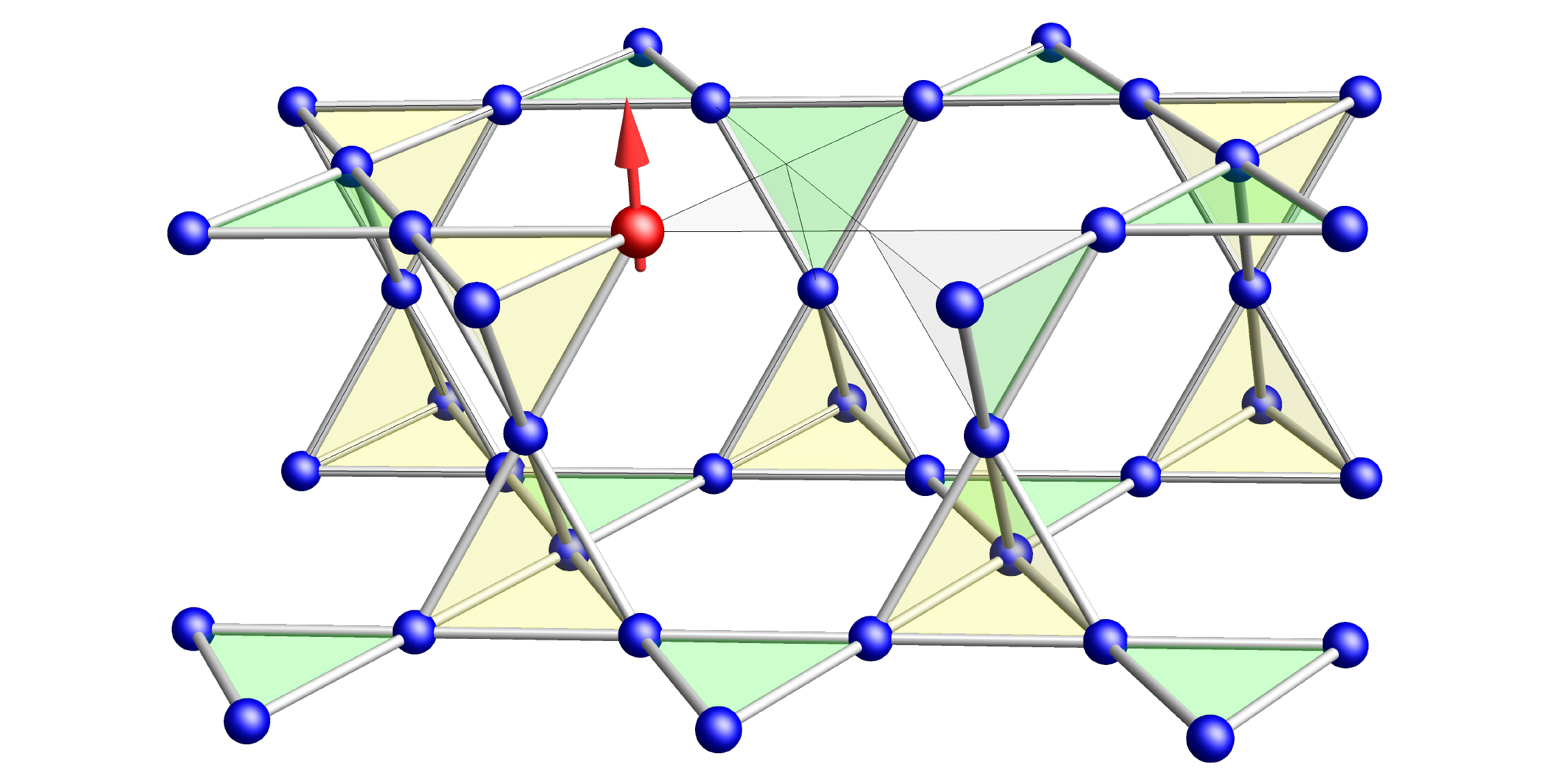}
\caption{
\label{fig:orphan_spin} (Color online) Orphan spin (red arrow) induced by vacancies in the kagome bilayer. An orphan spin resides in defective triangular simplex, in which two of the spins are removed, in either of the kagome layers.  From the viewpoint of simplex, two adjacent vacancies in a triangle removes one triangle simplex , but produces a $q=1$ (point) simplex, which is the orphan spin itself and transforming two neighboring tetrahedral  into triangular simplexes. 
}
\end{figure}

Since the presence of vacancies does not change the liquid nature or the frustrated spin-interactions in the kagome bilayer, it is unclear whether the non-magnetic impurities introduce any new dynamical effect. On the other hand, the so-called orphan spins due to the dilution are known to induce nontrivial effects on the equilibrium properties of the kagome bilayer~\cite{schiffer97}. The orphan spin corresponds to defect triangular simplex with only one surviving spin and two non-magnetic sites~\cite{henley01,moessner99}. An example of the orphan spin is shown in Fig.~\ref{fig:orphan_spin}. One can also think of orphan spin as connecting a $q=3$ triangular simplex and a $q=1$ point simplex, which is the spin itself.  The orphan spin behaves as a quasi-free spin with a fractionalized length $S/2$ when perturbed by a magnetic field~\cite{sen11,sen12}. 
Experimentally, these seemingly isolated free spins in diluted SCGO produce a Curie-like component in the static susceptibility even at temperatures well below $\Theta_{\rm CW}$~\cite{schiffer97,moessner99}.  Detailed Monte Carlo simulations uncover a complex spin texture surrounding the defect simplex whose total spins indeed sum to $S/2$~\cite{sen11,sen12}. The fractionalized spin-texture also provides a natural explanation for the short-range oscillating signal observed in nuclear magnetic resonance~\cite{limot02}.  

An intuitive argument for the fractionalized $S/2$ orphan spins was originally given by Henley from the viewpoint of bi-simplex structure~\cite{henley01}.
Because each spin is shared by two simplexes in the bi-simplex lattices such as the kagome-bilayer, the total magnetization can be written as 
$\mathbf M_{\rm tot} = \frac{1}{2} \sum_{\alpha} \mathbf L_\alpha$, where $\alpha$ now runs over tetrahedral, triangular, and $q=1$ simplexes in the presence of orphan spins. In the ground states, total spin of each tetrahedron and triangle simplex remains zero, as evidenced by Monte Carlo simulations~\cite{henley01}. As a result, the total magnetization of the system becomes $\mathbf M_{\rm tot} = \frac{1}{2} \sum_{\alpha}^{q=1} \mathbf L_\alpha$, where now the summation is restricted to $q=1$ single-point simplexes.  As shown above, such $q=1$ simplex is just the orphan spin itself, so we have $\mathbf M_{\rm tot} = \frac{1}{2} \sum_{i \in {\rm orphan}} \mathbf S_i$, which also means that each orphan spin can be viewed as a quasi-free spin with a fractionalized length~$S/2$ when in a magnetic field~\cite{henley01}.

\begin{figure}[t]
\includegraphics[width=0.95\columnwidth]{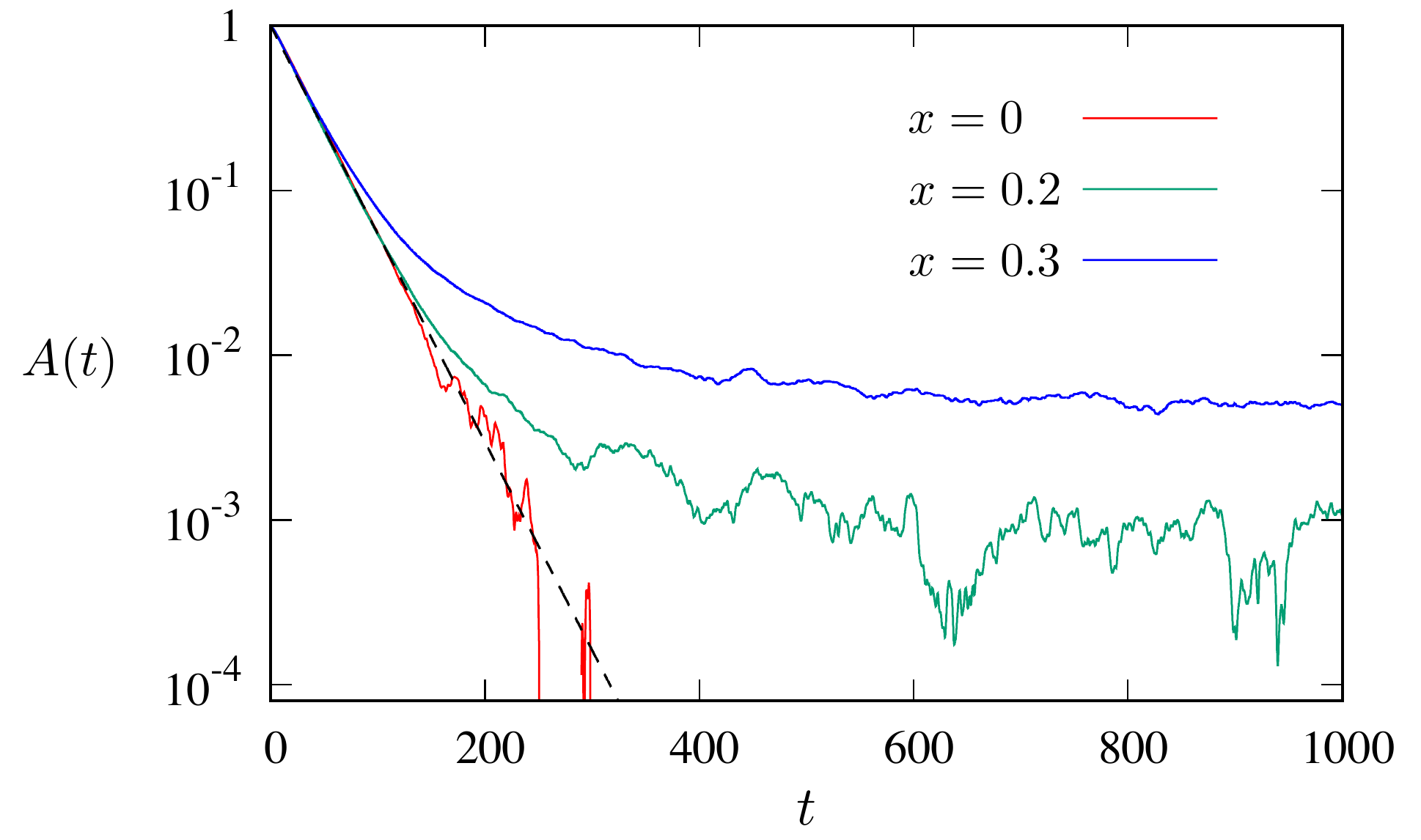}
\caption{  
\label{fig:At_cmp} (Color online) Semi-log plot of autocorrelation function $A(t)$ at $T = 0.01$ without vacancies (red) and with~20\% (green) and~30\% vacancies (blue), obtained from Landau-Lifshitz dynamics simulations of a $L = 48$ system. The black dashed line indicates an exponential decay $A(t) \approx e^{-t/\tau}$ at small $t$.  
}
\end{figure}

A natural question then is what is the dynamical manifestation of these orphan spins. To this end, we examine the spin-spin autocorrelation function $A(t)$ defined in Eq.~(\ref{eq:At}).  Fig.~\ref{fig:At_cmp} shows the semi-log plot of autocorrelation functions with and without vacancies obtained from our dynamical simulations. In both cases, the initial decay of the autocorrelation  can be well described by an exponential function, i.e. $A(t) \sim e^{-t/\tau}$ for small $t$.  However, while the exponential decay persists to longer time-scales in the non-diluted system, the autocorrelation function of the diluted magnet exhibits a long-time tail, indicating a significantly reduced decline rate of the spin-autocorrelation.  

This two-stage relaxation of the autocorrelation function can be understood in the framework of the two population picture~\cite{schiffer97} discussed previously, namely, the classical spin liquid of kagome bilayer can be viewed of consisting of the ``correlated" population which forms momentless clusters ($\mathbf L_\boxtimes = \mathbf L_\triangle = 0$) and the population of quasi-free ``orphan" spins that weakly interact with each other~\cite{sen12}. Of course, at very strong dilution, the set of free spins also include those completely isolated magnetic ions~\cite{henley01}. Dynamically, these two populations of spins are expected to behave differently.  As discussed in Sec.~\ref{sec:diffusion}, spin diffusion in the classical spin liquid, which is mainly driven by the zero energy modes, results in an autocorrelation function $A(t)$ which decays exponentially with time. On the other hand, since the vacancy-induced orphan spins can be viewed as nearly free spins, one expects their dynamical behavior to be similar to that of uncorrelated paramagnet. Earlier works have shown that diffusion of Heisenberg spins in an uncorrelated paramagnet leads to a power-law tail in the autocorrelation function, i.e. $A(t) \sim 1/t^\alpha$~\cite{muller88,gerling90,gerling90b,constantoudis97,bagchi13}, where the exponent $\alpha$ depends strongly on the dimensionality. For 2D Heisenberg magnet, it is estimated to be $\alpha \approx 1.05 \pm 0.025$~\cite{muller88}.

\begin{figure}[t]
\includegraphics[width=0.95\columnwidth]{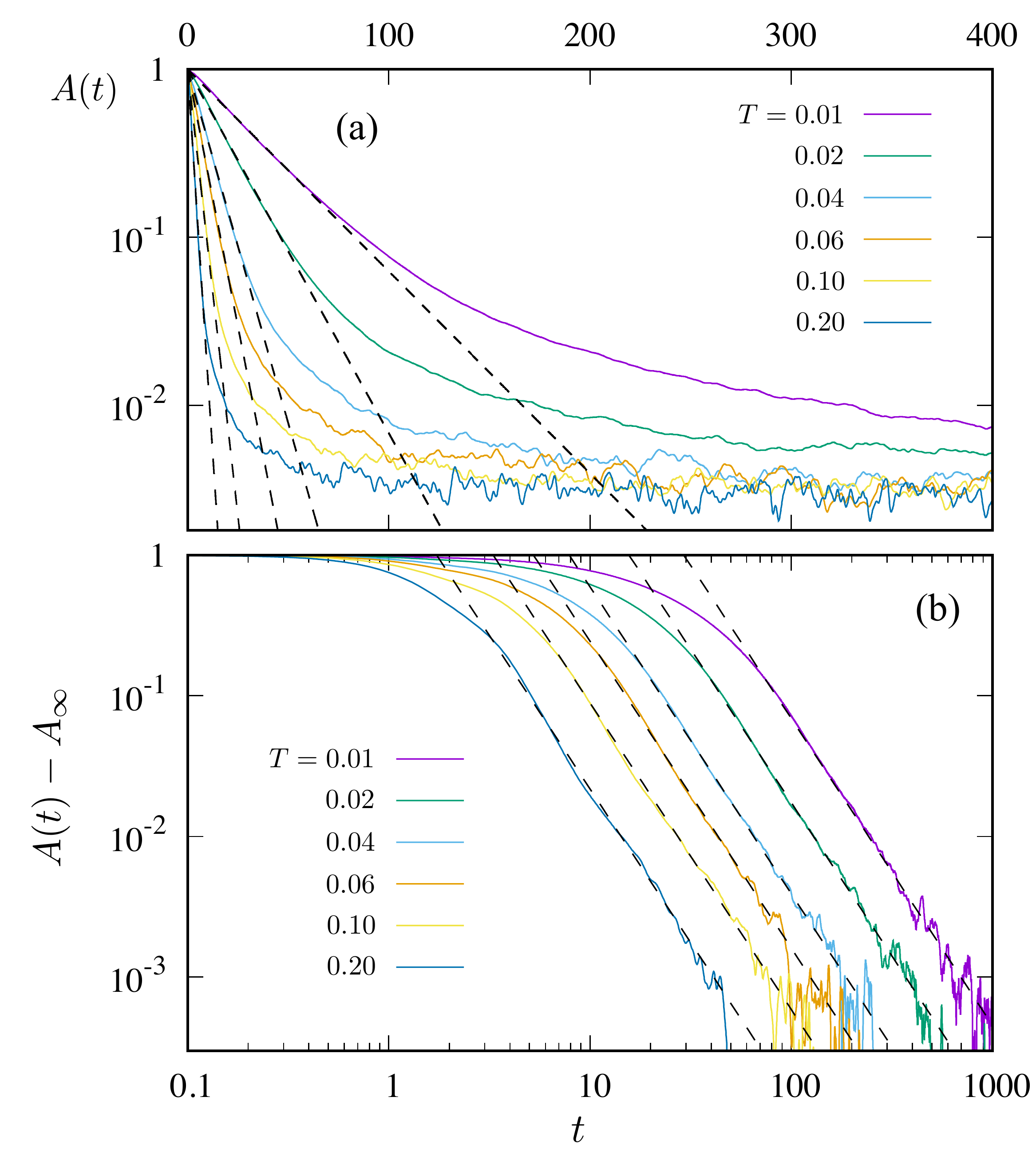}
\caption{  
\label{fig:At2} (Color online) (a) Semi-log plot of the spin autocorrelation function $A(t) = \sum_i \langle \mathbf S_i(t) \cdot \mathbf S_i(0) \rangle / N$ of $L = 48$ kagome-bilayer with~30\% vacancy at varying temperatures. The dashed lines correspond to the initial exponential decay of the autocorrelation function, i.e. $A(t) \sim \exp(-t/\tau)$ for small $t$. (b)~The log-log plot of same autocorrelation functions, with the asymptotic value at large time subtracted, at varying temperatures. The dashed lines indicate power-law long-tails $A(t) \sim \mathcal{C} / t^\alpha$, with an exponent $\alpha = 2.18$. 
}
\end{figure}

\begin{figure}
\includegraphics[width=.99\columnwidth]{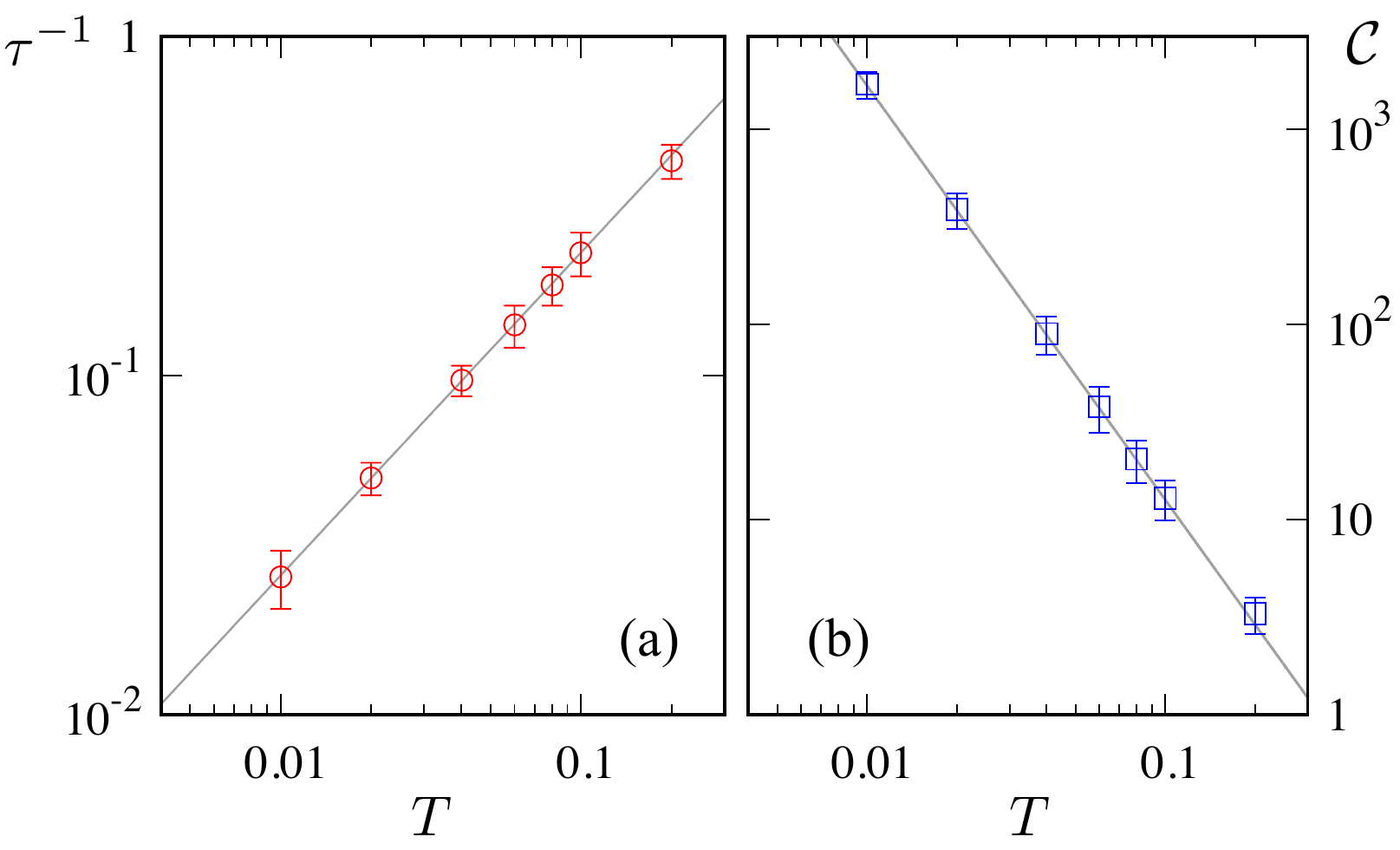}
\caption{   
\label{fig:corr2-param} (Color online) (a) The decay-time constant $\tau$ as a function of temperature $T$ in log-log plot. The solid line corresponds to a power-law dependence $\tau \sim 1/T^{0.953}$. (b) The amplitude $\mathcal{C}$ of the power-law long-tail versus the temperature. The straight line of the log-log plot indicates a power-law relationship $\mathcal{C} \sim 1/T^{2.123}$.
}
\end{figure}

To verify the above picture, we present detailed examination of both the short-time and long-time behaviors of spin autocorrelation function for a diluted kagome bilayer with a vacancy density $x = 0.3$. We performed extensive Monte Carlo and Landau-Lifshitz dynamics simulations on a $L = 48$ lattice (with total number of spins $N = 7 \times L^2 = 16128$). Over 50 independent realizations of the disorder were constructed, and for each vacancy configuration, 100 independent initial spin states are prepared at the simulation temperatures. Fig.~\ref{fig:At2}(a) shows semi-log plot of spin-autocorrelation at various temperatures. At short time scale, the decrease of $A(t)$ can be reasonably approximated by an exponential decay $A(t) \sim e^{-t/\tau(T)}$, similar to the undiluted case, with a temperature-dependent decay time constant $\tau$. The numerically extracted relaxation time $\tau$, shown in Fig.~\ref{fig:corr2-param}(a), again exhibits a power-law dependence on temperature $\tau \sim 1/T^\zeta$, with an exponent~$\zeta \sim 0.952 \pm 0.017$, which is similar to the undiluted case. As discussed in the previous section, the spherical approximation for the classical spin liquid predicts an exponent $\zeta = 1$. It is unclear whether the deviation here is due to finite-size effect or the soft-spin approximation.

At longer time scales, the decay of the autocorrelation function slows down and turns into a power-law tail, $A(t) \sim   A_\infty + \mathcal{C}(T)/t^\alpha $, with the same exponent $\alpha$ for different temperatures. Interestingly, as shown in Fig.~\ref{fig:corr2-param}(b), the amplitude of this power-law tail also exhibits a power-law dependence $\mathcal{C} \sim 1/T^{\eta}$, with an exponent $\eta = 2.123 \pm 0.021$. It is also worth noting that the decay of spin-autocorrelation saturates to a small but non-zero value at large times, as shown in Fig.~\ref{fig:corr2-param}(a). Similar results, which can be attributed to finite-size effect, have been reported in the spin-dynamics of uncorrelated classical Heisenberg chains~\cite{constantoudis97,bagchi13}. 

To better understand the power-law decay and the origin of the nonzero asymptotic $A_\infty$ in the diluted systems, we consider the dynamics of an orphan spin. At any finite temperatures,  the total spin of individual simplex does not vanish identically, hence the ground-state condition Eq.~(\ref{eq:gs}) is not strictly satisfied. Indeed, the fluctuation of simplex magnetization is given by $\langle \mathbf L_\alpha^2 \rangle \sim T/J$~\cite{moessner98a,sen12}. This also indicates a non-zero coupling between orphan spins and the background correlated spin liquid. This residual coupling leads to incoherent precession of orphan spins and the exponential decay of the orphan-spin autocorrelation function. On the other hand, as shown in Ref.~\cite{sen12}, there is an emergent effective interaction between the orphan spins. At low enough temperatures, their collective dynamics induced by this residual interaction thus slows down the exponential decay of autocorrelation function that is caused by coupling to the background spin liquid, and turns it into a power-law decay, similar to the anomalous spin diffusion in classical Heisenberg magnets at high temperatures~\cite{muller88}.

Interestingly,  the exponent $\alpha \approx 2.1$ obtained from our numerical fitting is significantly different from that of the 2D paramagnet. This unusual result could be attributed to the complex interaction between the orphan spins. As demonstrated in Ref.~\onlinecite{sen12}, there is an emergent Heisenberg exchange interaction between the orphan spins, that is determined by the charge-charge correlator of the underlying Coulomb spin liquid. Moreover, the sign (FM vs AFM) depends on whether the two orphan spins belong to the same kagome layer or not. It has been speculated whether this complex and potentially frustrated interaction might lead to glassy dynamics at low temperatures. Indeed, although it is believed that the autocorrelation function of spin-glass exhibits an stretched exponential decay at temperatures above the glass transition $T_g$, general scaling rules near the glass transition point imply a cutoff power-law~\cite{palmer84,sompolinsky82,ogielski85,pinkvos90,keren96}, such as the Ogielski form $A(t) \sim t^{-\alpha} \, \exp[-(\lambda t)^\beta]$, where the parameter $\lambda \to 0$ as $T \to T_g$. If the kagome-bilayer can be viewed as exhibiting a glass transition at $T = 0$, as conventional 2D spin-glasses, the autocorrelation function might be dominated by a power-law behavior at intermediate time scale before it is cut off by the stretched exponential. Further larger scale simulations are required to investigate this scenario.


The power-law tail and the associated collective behaviors also depend strongly on the density $x$ of orphan spins.  The effective interaction between two such defects separated by a distance $ r$ is given by $J_{\rm eff}(r) \sim  T \mathcal{J}( r / \xi(T))$, where $\xi \sim 1/\sqrt{T}$ is the temperature-dependent correlation length of the background spin liquid, and the function   $\mathcal{J}(r) \sim \exp(-r)$  decays exponentially at large distances~\cite{sen12}. The $T$-linear prefactor here indicates the entropic origin of the effective interaction, namely $J_{\rm eff}$ arises from conformational entropy of the fluctuating background correlated spins. Since the average distance between orphan spins scales as $\ell \sim 1/\sqrt{x}$, one thus obtains an average  interaction $\overline{J}_{\rm eff} \sim J_{\rm eff}(\ell)  \sim T\, \exp(-\sqrt{T/x})$, which becomes exponentially weak at small vacancy percentages. Despite this weakened interaction, the collective behavior of orphan spins would set in at a temperature that is of the order of the effective interaction. This gives the condition: $T^* \sim \, \overline{J}_{\rm eff}(T^*)$. Using the expression for $\overline{J}_{\rm eff}$ above, one thus obtain a characteristic temperature $T^* \sim x$ that decreases linearly with the reduced defect density. Physically, the thermal correlation length at this $T^*$ is comparable to inter-orphan-spin distance.



\section{Discussion and outlook}
\label{sec:discussion}

To summarize, we have extensively characterized the spin dynamics in the liquid phase of Heisenberg antiferromagnet on the kagome bilayer, which is relevant for the frustrated magnet SCGO. By computing the dynamical structure factor at different temperatures and dilutions, we show that the spin excitations are dominated by spin diffusion in the low energy, long time regime. The spin diffusion constant depends weakly on temperature, but decreases with dilution. Another interesting result is the half moon pattern of the dynamical structure factor with energy $\omega \gtrsim J$. Similar features have recently been observed in some pyrochlore compounds, it remains to be seen whether these remnants of the propagating spin waves can be observed in SCGO.
Our simulations on diluted bilayer kagome shows that spin diffusion remains the dominant process in the presence of site disorder. This result further confirms, from the dynamical viewpoint, that site-disorder itself does not immediately cause glassy behaviors in the classical spin liquid, although the diffusion relaxation time becomes longer with increasing disorder. However, for disorder due to non-magnetic vacancies, the presence of so-called orphan spins results in an intriguing power-law tail in the spin-autocorrelation function. This power-law slow dynamics indicates that the system might be on the verge of a glass transition, which could be induced by other perturbations.


As discussed above, our work offers important benchmark for future dynamics studies of kagome bilayer that include other perturbations. Of particular interest are those perturbations that might transform the classical spin liquid into either the conventional spin glass or the more exotic spin jam. Indeed, since the diffusive spin dynamics in highly frustrated magnets is mainly driven by the zero-energy modes, one expects a diminishing diffusivity when the number of such zero modes is significantly reduced. For example, the entropic barrier in the coplanar phase of kagome reduces the {\em continuous} weather-van modes to {\em discrete} zero modes defined on closed loops. It has been proposed that the much slower relaxation of these discrete loops might give rise to glassiness without intrinsic disorder in kagome~\cite{cepas12,cepas14}.  However, the coplanar phase induced by thermal order-by-disorder seems to remain a classical spin liquid~\cite{taillefumier14}. A transition into the glassy regime might still occur at a lower temperature when the dynamics is dominated by quantum tunneling of loops~\cite{cepas12}.

Contrary to kagome Heisenberg antiferromagnets, there is no thermal induced coplanar or collinear phase in kagome bilayer. On the other hand, it has been proposed by one of us and co-authors in Ref.~\cite{klich14} that a coplanar regime, in which spins in each tetrahedron are collinear, can be induced by quantum fluctuations. Moreover, different coplanar ground states can be mapped to discrete hexagonal tilings. Importantly, there is no continuous weather-van modes in this coplanar regime, and the only zero-energy modes are system-wide extended strings~\cite{klich14}. As jamming transition often occurs in such constrained discrete models, the resultant coplanar phase is dubbed the spin jam~\cite{klich14,yang15}. It is argued that quantum fluctuations transform the degenerate classical ground-state manifold into a rugged landscape that is different from that of conventional spin glass. While this spin-jam picture seem to explain some properties of SCGO and other similar glassy magnets, such as the much weaker memory effect~\cite{samarakoon16,samarakoon17}, an important open question is to see how dynamical behaviors characteristic to spin-jam evolve from the classical spin liquid, which will be left for future study.


\bigskip

\begin{acknowledgements}
We thank Anjana Samarakoon and Israel Klich for useful discussions and for collaborations on related projects. DZ and SHL were supported by the U.S. Department of Energy, Office of Science, Office of Basic Energy Sciences under Award Number DE-SC0016144. 
The author also acknowledge the support of Advanced Research Computing Services at the University of Virginia.
\end{acknowledgements}

\end{document}